\documentclass[aps,prd,twocolumn,showpacs,nofootinbib]{revtex4}
\usepackage{graphics}  % needed for figures
\pdfoutput=1
\usepackage{epsfig}
 \usepackage{verbatim}

\IfFileExists{srcltx.sty}{\usepackage[active]{srcltx}}

\newcommand{\sun}{\ensuremath{\odot}}%                          % sun symbol
\newcommand{\be}{\begin{equation}}
\newcommand{\ee}{\end{equation}}
\newcommand{\bea}{\begin{eqnarray}}
\newcommand{\eea}{\end{eqnarray}}

\begin{document}
%\vskip 20cm

\title{Astrophage of neutron stars from supersymmetric dark matter Q-balls}

\author{Ian M. Shoemaker}
\affiliation{Department of Physics and Astronomy, University of California, Los Angeles, CA 90095-1547, USA}
% \date{\today}
%\preprint{UCLA/09/TEP/??}

\begin{abstract}
The gauge-mediated model of supersymmetry breaking implies that stable non-topological solitons, Q-balls, could form in the early universe and comprise the dark matter. It is shown that the inclusion of the effects from gravity-mediation set an upper limit on the size of Q-balls. When in a dense baryonic environment Q-balls grow until reaching this limiting size at which point they fragment into two equal-sized Q-balls. This Q-splitting process will rapidly destroy a neutron star that absorbs even one Q-ball. The new limits on Q-ball dark matter require an ultralight gravitino $m_{3/2} \lesssim \rm{keV}$, naturally avoiding the gravitino overclosure problem, and providing the MSSM with a dark matter candidate where gravitino dark matter is not viable. 
\end{abstract}

\maketitle
%%%
\section{Introduction}
%%%
Supersymmetry is a plausible candidate for physics beyond the standard model. All supersymmetric extensions of the standard model include non-topological solitons, or Q-balls, in their spectra~\cite{Kusenko:1997zq}. In the minimal supersymmetric standard model (MSSM) it is the flat directions which allow for the existence of Q-balls.  In the early universe, large Q-balls are abundantly produced from the fragmentation of a flat direction condensate. If gauge-mediated supersymmetry breaking is present in nature, these Q-balls can be absolutely stable~\cite{Dvali:1997qv} and exist today as dark matter~\cite{Kusenko:1997si}. The cosmology and astrophysical implications of Q-balls have been studied by a number of authors~\cite{Frieman:1988ut,Frieman:1989bx,Griest:1989bq,Griest:1989cb,Enqvist:1997si,Enqvist:1998ds,Enqvist:1998xd,Kusenko:1997ad,Kusenko:1997hj,Kusenko:1997it,Kusenko:1997vp,Laine:1998rg,Enqvist:1998en,Enqvist:1998pf,Axenides:1999hs,Banerjee:2000mb,Battye:2000qj,Allahverdi:2002vy,Enqvist:2003gh,Dine:2003ax,Kusenko:2004yw,Kusenko:2005du,Berkooz:2005rn,Berkooz:2005sf,Kusenko:2008zm,Johnson:2008se,Kasuya:2008xp,Sakai:2007ft,Campanelli:2007um,Kasuya:2000wx,Kawasaki:2005xc,Kasuya:2007cy,Shoemaker:2008gs,Campanelli:2009su,Kusenko:2009iz}. 

Even within the gauge-mediated paradigm however, the effects from gravity-mediation come to dominate at sufficiently large field amplitude. These gravity contributions completely alter Q-ball properties~\cite{Kasuya:2000sc,Kasuya:2005ay}. We will assume the most common paradigm for gauge-mediation in which there is a hidden supersymmetry breaking sector, a messenger sector and the visible MSSM sector, all coupled via weak gauge interactions \cite{PhysRevD.48.1277,PhysRevD.51.1362,PhysRevD.53.2658}.  The effects of gravity are included by embedding the theory in minimal supergravity as done in~\cite{deGouvea:1997tn,Giudice:1998bp}. The most studied particle dark matter candidate in gauge-mediated models is the gravitino. However for the gravitino LSP to be dark matter there is a lower bound on the mass $m_{3/2} \gtrsim 100~ \rm{ keV}$ ~\cite{Steffen:2006hw}. Thus the gauge-mediated scenario is devoid of a particle dark matter candidate for very low gravitino mass. 

We show that the inclusion of gravity-mediation effects causes Q-balls above a critical size to split into two equal-sized daughter Q-balls. This immediately  eliminates the ``new-type'' Q-balls~\cite{Kasuya:2000sc} as dark matter candidates. Moreover, any neutron star that encounters even one Q-ball in its lifetime will be rapidly consumed by the exponential growth of Q-balls coming from this splitting process. For Q-balls to be dark matter, neutron star lifetimes require $m_{3/2} \lesssim \rm{keV}$, giving the MSSM a natural dark matter candidate in a regime where gravitino dark matter is not viable.  Such a light gravitino mass naturally avoids the gravitino overclosure problem without any constraint on the reheating temperature~\cite{Giudice:1998bp}.  Note however that it has previously been pointed out that the existence of flat-directions in supersymmetric theories may by itself solve the gravitino problem through late thermalization after delayed inflaton decay \cite{Allahverdi:2007zz}, thereby substantially lowering the reheating temperature \cite{Allahverdi:2005mz}. 

The remainder of the paper is organized as follows. In Section I. we show that the effects of gravity alter Q-ball properties, which forces a sufficiently large Q-ball to split in two. In Section II. we consider the effects Q-balls on neutron star lifetimes and deduce stringent new bounds on Q-ball dark matter and the gravitino mass. We conclude in Section IV. 

%%%
\section{The Q-split}
%%%
Although we assume a gauge-mediated model of supersymmetry breaking, the effects of gravity cannot be neglected at sufficiently large VEV. Radiative corrections to the flat directions of the MSSM dictate the form of the potential~\cite{Kusenko:1997si, Enqvist:1997si,Kasuya:2000sc}
\bea V(\phi) &=& M_{s}^{4} \log \left( 1 + \frac{| \phi|^{2}}{M_{s}^{2}}\right) \nonumber \\ 
&+& m_{3/2}^{2} |\phi|^{2} \left[ 1 + K \log \left( \frac{|\phi|^{2}}{M^{2}}\right)\right], \label{pot} \eea
where $m_{3/2} $ is the gravitino mass, $K$ is constant coming from one-loop corrections, and $M_{s}$ is the supersymmetry breaking scale. The properties of the Q-ball change dramatically when $\phi$ becomes so large that gravity effects are comparable or dominate. At small VEV only the gauge-mediated effects are relevant and the second term in Eq. \ref{pot} can be ignored. In this regime the Q-ball is of the gauge-mediated type with mass $M_{G}(Q) \sim M_{s} Q^{3/4}$. To be stable a Q-ball must not decay into fermions, scalars or other solitons. The Q-ball is by construction the state of minimum energy with respect to the scalars.  The stability of Q-balls with respect to fermionic modes of decay depends on their baryon number. So long as $ Q \gtrsim Q_{st} \equiv \left(\frac{M_{s}}{1~ \rm{GeV}}\right)^{4} \sim 10^{12}$ the decay into fermionic baryons will be kinematically forbidden.  The final check on stability should be with respect to the solitonic sector. We must check that the mass of a single Q-ball of charge $Q$ is less than the mass of two Q-balls each with charge $Q/2$.  For Q-balls of the gauge-mediated type (when only the first term in Eq. \ref{pot} is relevant) it is the fractional power of the mass-charge relation that guarantees this for all $Q$.  Thus the gauge-type Q-balls are rendered completely stable for sufficiently large $Q$. 

The new-type Q-balls are stable with respect to bosonic and fermionic modes of decay~\cite{Kasuya:2000sc}, but not with respect to decay in the solitonic sector. Gravity effects become important when $\phi \gtrsim \phi_{eq} \equiv M_{s}^{2}/m_{3/2}$. At such a point the mass of the Q-ball becomes
\be M_{H} (Q)  \sim m_{3/2} ~Q. \ee
We can turn a gauge-type Q-ball into one of the new-type by dumping enough charge into it, such that the VEV becomes comparable to $\phi_{eq}$.   Note the the VEV of the gauge-type Q-ball depends on the charge as $\phi \sim M_{s} Q^{1/4}$. In other words the critical value of $Q$ for which new-type Q-balls form is 
\be Q_{split} \sim \left(\frac{M_{s}}{m_{3/2}}\right)^{4}. \label{qsplit} \ee  
It is apparent then that $M_{H}(2 Q_{split}) \gtrsim 2 M_{G} (Q_{split})$, which implies that all hybrid-type Q-balls immediately split into two equal sized gauge type Q-balls. Thus the hybrid-type Q-balls as envisaged in~\cite{Kasuya:2000sc} cannot be dark matter since they immediately fragment into Q-balls of the gauge-mediated type. 

The Q-split has important ramifications for the constraints on gauge-mediated type Q-balls as well, requiring $Q \lesssim Q_{split}$. Even in the absence of the astrophysical limits that follow, the existing Super-K limit on $Q$ requires $Q \gtrsim 10^{24}$ which already imposes the interesting limit $m_{3/2} \lesssim 10^{-3}~ \rm{GeV}$. 
%%%%%
\begin{figure}
\begin{center}
\includegraphics[scale=0.45]{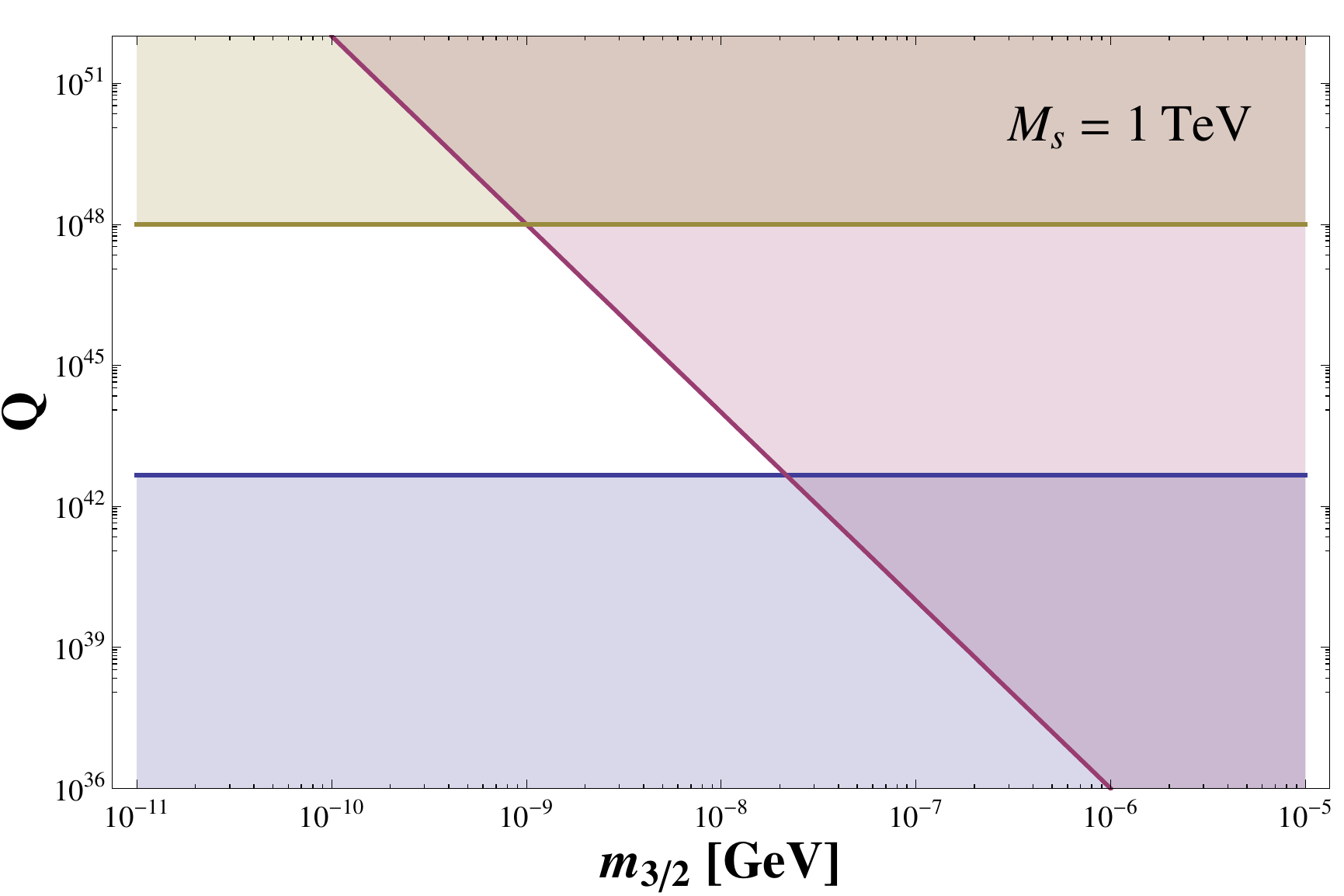}
\caption{The bottom half of the plot is excluded because these Q-balls would destroy neutron stars too rapidly (See Eq. \ref{cap}).  The top half of the plot is excluded by Eq. \ref{mmin}. The diagonal upper-right slice is excluded by the Q-split process, Eq. \ref{qsplit}.}
\end{center}
\label{figure1}
\end{figure}
%%%%%%

%%%%%
\section{Astrophage from the Q-split}
%%%%%
While although ordinary stars and planets are insufficient to stop Q-balls that pass through them, a neutron star can~\cite{Kusenko:1997vp}. The physics of a Q-ball in a neutron star has been studied before~\cite{Kusenko:1997it,Kusenko:2005du}.  Here however we include the dramatic effects of the Q-split. The authors~\cite{Kusenko:2005du} showed that phenomenologically acceptable Q-balls must have higher dimensional operators which violate $U(1)_{B}$ symmetry and thus limit Q-ball size to be below a certain $Q_{cr}$ depending on the flat direction. In the remainder of the paper we assume a flat direction for which $Q_{split} \lesssim Q_{cr}$ such that we may ignore the effects of higher dimensional operators. 

Once a neutron star captures its first Q-ball, a dramatic transition occurs in a short time period.  After stopping in the neutron star the size of the Q-ball grows as it imbibes neutrons. The Q-ball eventually splits when $Q \sim Q_{sp}$ and the daughter Q-ball themselves grow to split further. Note that in contrast with previous studies~\cite{Kusenko:1997it,Kusenko:2005du} the final state is neither a giant Q-ball nor a black hole, but rather a gravitationally bound ball of Q-splits. 

The growth of the Q-ball towards the Q-split solution is complicated by the fact that the $Q$ dependence of the Q-ball radius interpolates between $R_{gauge} \approx M_{S}^{-1} Q^{-1/4}$ and $R_{grav} \approx |K|^{-1/2} m_{3/2}^{-1}$, as it transitions to from a gauge-type to gravity-type solution~\cite{Enqvist:1997si,Enqvist:1998en,Enqvist:1998ds,Kasuya:2000sc}. We expect the difference between these two solutions to be maximal when $R_{grav} \sim R_{gauge}$, which occurs at $Q_{R} \sim |K|^{-2} Q_{sp}$. Since $|K| \lesssim 0.1$ we expect $R_{gauge}$ to be a good approximation of the true Q-ball radius even as it approaches the Q-split.   

Assuming that the growth rate is the same as the rate at which baryons fall on the Q-ball surface, we have
\be \frac{dQ}{dt} = 4 \pi R_{Q}^{2} n_{ns} v \sim \left(\frac{4 \pi n_{ns}v}{M_{s}^{2}} \right) Q^{1/2} \equiv \alpha~\frac{ Q^{1/2}}{M_{s}^{2}} \ee
where $n_{ns} \sim 10^{-3}~ \rm{GeV}^{3} $, $v \sim 10^{-3}$, and $\alpha \sim 10^{-5}~ \rm{GeV^{3}}$.  However it was argued in~\cite{Kusenko:2005du} that realistically one must include the fact that a Q-ball converts nucleons to their respective antiparticles~\cite{Kusenko:2004yw}, the annihilations of which create a large pion pressure in the vicinity of the Q-ball. In this case the rate of baryon absorption is determined by the relative pion and neutron pressures. This is equivalent to substituting $\alpha$ in the above for the hydrodynamic corrected $\alpha' \sim 10^{-7}~ \rm{GeV}^{3}$. 

After a time $t_{split}$ the Q-ball of initial size $Q_{0}$ will have grown to a critical size for the Q-split to form:
\be t_{split} = \frac{2}{\alpha'} \left[ Q_{split}^{1/2} - Q_{0}^{1/2}\right]. \ee 
When $Q_{0} \ll Q_{split}$ the time until the first Q-splitting occurs is independent of the initial charge of the Q-ball. For neutron star densities the time scale is quite short 
\be t_{split} \sim 10^{-5} \left( \frac{M_{s}}{\rm{TeV}}\right)^{4} \left( \frac{\rm{GeV}}{m_{3/2}}\right)^{2} \rm{s}. \ee
Once the Q-split forms inside the neutron star its fate is determined. From this point forward the number of Q-balls grows exponentially. A Q-split forms two smaller Q-balls each of charge $Q_{split}/2$. Then each of these grows in a time roughly $t_{split}$ to fragment again. This process continues until all the baryonic fermions have been converted into squarks. We call this process astrophage\footnote[2]{From the Greek, $\acute{\alpha} \sigma \tau \rho o \nu$ = star, and $\phi \acute{\alpha} \gamma \epsilon \iota \nu$ = to eat.} since the Q-ball will consume the entire neutron star. The baryon number inside the Q-balls of the neutron star grows as 
\be N_{Q}(t) \sim Q_{split}~ 2^{t/t_{split}}. \ee 
As a simple example consider a Q-ball with parameters $Q\sim 10^{24}$, $M_{s} \sim 1~ \rm{TeV}$, and $m_{3/2} \sim 10^{-4}~\rm{GeV}$. In this case every neutron inside the star is converted to squarks in a time 
\be t_{*} \sim \frac{t_{split}}{\log 2} ~  \log \left( \frac{Q_{NS}}{Q_{split}}\right) \lesssim 10^{2}~ t_{split}. \ee
Combining this with our expression for the Q-splitting timescale, we see that lifetime of a neutron star once astrophage starts is
\be t_{*} \lesssim 10^{9}~   \left( \frac{M_{s}}{\rm{TeV}}\right)^{4} \left( \frac{\rm{keV}}{m_{3/2}}\right)^{2} \rm{s}. \ee

The limiting factor in this analysis is the Q-ball capture rate of a neutron star. For sufficiently large Q-balls the number density is so low that none ever encounter a neutron star in a cosmologically relevant time scale. The flux of dark matter Q-balls is 
\be F_{Q} \sim \frac{v~ \rho_{DM} }{4 \pi M_{Q}} \sim \frac{10^{2}}{Q^{3/4}} \left(\frac{\rm{TeV}}{M_{s}}\right)~ \rm{cm}^{-2} \rm{s}^{-1} \rm{sr}^{-1}, \ee
where $\rho_{DM} = 0.3 ~\rm{GeV}/\rm{cm}^{3}$ and $v\sim 10^{-3}$. 
The time for a neutron star to capture one Q-ball is roughly 
\be \tau_{cap} \sim \frac{1}{4 \pi R_{NS}^{2} F_{Q}}, \ee
where a typical neutron star radius is $R_{NS} \sim 10~ \rm{km}$. The age of the oldest neutron stars come from millisecond pulsars which typically have a characteristic age of order $10~ \rm{Gyr}$ ~\cite{Hansen:1997ha, Hansen:1997hb, lrr-2001-5}.  Phenomenologically acceptable Q-balls must have $\tau_{cap} > 10^{10}~\rm{yr}$. We can reexpress this is a limit on Q-ball size: 
\be Q \gtrsim 10^{43}~\left(\frac{\rm{TeV}}{M_{s}}\right)^{4/3}. \label{cap} \ee

Thus Q-balls that are sufficiently heavy have such low fluxes that they have never been captured by a neutron star and are therefore not excluded by the astrophage process. Thus the allowed Q-balls are those that satisfy 
\be 10^{43} \left(\frac{\rm{TeV}}{M_{s}}\right)^{4/3} \lesssim Q \lesssim Q_{split}. \ee

This can be turned into a upper bound on the gravitino mass for phenomenologically acceptable Q-ball dark matter
\be \left(\frac{m_{3/2}}{\rm{keV}}\right) \lesssim 10^{-2}~ \left(\frac{M_{s}}{\rm{TeV}}\right)^{4/3}. \ee

\section{New constraints on Q-ball dark matter}

We summarize the new constraints on Q-ball properties in Fig. 1 for $M_{s} \sim 1~ \rm{TeV}$. In this plot the upper limit on Q-ball size is given by the $Q_{split}$ size in Eq. \ref{qsplit}. All Q-balls below a certain size are captured by neutron stars and therefore excluded, Eq. \ref{cap}. Altogether the viability of Q-ball dark matter tightly constrains the gravitino mass and the minimum Q-ball baryon number (see Table \ref{m32 limits}). With an ultralight gravitino there is no bound on the reheating temperature and the gravitino problem is naturally avoided~\cite{Giudice:1998bp,deGouvea:1997tn}. The existence of an ultralight gravitino can be discovered at the LHC~\cite{Hamaguchi:2007ji,Shirai:2009kn} and provide compelling evidence for the above scenario. The existence of Q-balls for such low gravitino mass is beneficial, since thermal gravitino dark matter requires $m_{3/2} \gtrsim \rm{keV}$, yet Q-ball dark matter is viable when $m_{3/2} \lesssim \rm{keV}$.  Moreover in this scenario the new lower limit on the baryon charge $Q \gtrsim 10^{43}$ is too large for the baryogenesis through partial Q-ball evaporation as envisioned in~\cite{Laine:1998rg,Banerjee:2000mb}.  

The production of such large Q-balls from the primordial condensate fragmentation is not difficult \cite{Enqvist:2000cq,Kasuya:2001hg}. Kasuya and Kawasaki have performed 3D lattice simulations of Q-ball formation in gauge-mediation and shown that the simple scaling relation for the largest charge formed from AD fragmentation holds quite well $ Q\sim 10^{-3} \left(\phi_{0}/M_{s}\right)^{4}$. To form  $Q \gtrsim 10^{43}$ Q-balls thus requires $\phi_{0} \gtrsim 10^{12}~ \rm{GeV}$, which is a modest constraint since large amplitude in AD models is not difficult to obtain~\cite{Dine:2003ax,Kasuya:2001hg}. Note that subplanckian field amplitudes imply that the charge can be as large as $Q \lesssim 10^{57}$.  Though such large charges are possible initially, these super-Q-split states will rapidly decay into a large number of gauge-type Q-balls. 

All bounds on Q are given in Table \ref{m32 limits} as a function of the supersymmetry breaking scale $M_{s}$. We report the phenomenological bounds $Q_{min} \lesssim Q \lesssim Q_{max}$, using $Q_{min} \equiv \rm{max}$$~(Q_{exp},Q_{ast})$ and $Q_{max} \equiv Q_{sp}$, where $Q_{exp}$ is the lower limit on Q from direct search experiments such as Super-K~\cite{Arafune:2000yv, Takenaga:2006nr}, $Q_{ast}$ is the limit (Eq.~\ref{cap}) that avoids astrophage, and $Q_{sp}$ is (Eq.~\ref{qsplit}) the charge at which a Q-split is formed. Table~\ref{m32 limits} also includes $m_{32}^{max}$, the maximum allowed gravitino mass consistent with the astrophage constraint on Q-ball dark matter. This makes the improvement over direct search experiments apparent as the astrophage limits represent a nearly 20 order of magnitude improvement in the bounds on $Q$. The upper bound $Q_{max}$ will be of order $Q_{sp}$ when $m_{3/2}$ takes the smallest possible value. In gauge-mediation the gravitino mass is generally $m_{3/2} \sim \Lambda_{DSB}^{2}/M_{P}$, where $\Lambda_{DSB}$ is the scale at which the gauge interactions in the SUSY-breaking sector become strong.  In such a scenario the gravitino mass may be as low as $\mathcal{O}(\rm{eV})$. In this limiting case the critical Q-split charge can be as large as 
\be Q_{max} \sim 10^{48}~\left(M_{s}/\rm{TeV}\right)^{4}. \label{mmin} \ee 
From Table \ref{m32 limits} one can see that the bounds get progressively weaker as the supersymmetry breaking scale increases. 

\begin{table}[htdp]
\caption{Bounds on charge Q for Q-ball dark matter as a function of supersymmetry breaking scale. Each limit also places an upper limit on the gravitino mass. The most up-to-date experimental bounds on $Q$ are reported in \cite{Arafune:2000yv}.}
\begin{center}
\begin{tabular}{|c|c|c|c|c|c|}
\hline
$~M_{s}~$  & $~m_{3/2}^{max}~ $ & $~Q_{exp}~$& $~Q_{ast}~$& $~Q_{min}~$& $~Q_{max}~$   \\ 
\hline
$1~ \rm{TeV}$ & $10~ \rm{eV}$ & $10^{24}$ & $10^{43}$&$10^{43}$&$10^{48}$\\
$10~ \rm{TeV}$ & $200~ \rm{eV}$& $10^{23}$&$10^{42} $&$10^{42} $& $10^{52}$\\
$100~ \rm{TeV}$ & $5~\rm{keV}$& $10^{22}$&$10^{40} $&$10^{40} $ & $10^{56}$\\
\hline
\end{tabular}
\end{center}
\label{m32 limits}
\end{table}

One can weaken the constraints on Q-ball dark matter by revoking our original assumption of embedding the MSSM in minimal supergravity. If nature favors no-scale supergravity or supergravity with Heisenberg symmetry then the gravity correction to the flat direction potential does not exist~\cite{deGouvea:1997tn}.  In this case then the gauge-type Q-ball solution persists to arbitrarily large field values and the astrophage of neutron stars never occurs. The best constraints on such Q-balls come from Super-K and MACRO~\cite{Kusenko:1997vp, Arafune:2000yv, Takenaga:2006nr}. The best prospect for detecting (or improving experimental bounds on) these states may come from the anomalous neutrino flux produced from the terrestrial passage of such Q-balls.~\cite{Kusenko:2009iz}.

Lastly, let us comment on the observational consequences of a destabilized neutron star.  It has been previously suggested that a similar process of Q-ball precipitated neutron star destabilization may account for some gamma-ray bursts \cite{Kusenko:1997it}.  While although nearly all of the phenomenologically viable parameter space for dark matter Q-balls excludes the possibility of such destabilization in our universe since all neutron stars would be destroyed, the flux of Q-balls may be low enough that only a fraction of neutron stars are destroyed. Two possibilities for such a scenario exist: dark matter Q-balls ($\Omega_{Q} \sim 0.2$) which are very close to the $Q\sim10^{43}$ limit for neutron star capture may not destroy all neutron stars but merely some fraction of them; or Q-balls for which $\Omega_{Q} \ll 1$ would also not be captured by neutron stars at too high a rate.  As the Q-splits rapidly consume the neutron star, the mass decreases below the minimum neutron star size $M_{min} \approx 0.2 M_{\sun}$, at which point the star explodes.  The newly free neutrons decay into protons, electrons and neutrinos which subsequently produce gamma-rays. The total energy released in such an explosion is roughly $10^{54} \rm{erg}$.  It has been previously noted \cite{Paczynski:1986px,Kusenko:1997it} that for such a scenario to account for gamma-ray bursts the dying neutron stars must be located at around $z = 1 -2$.  Though the precise details of such an explosion are beyond the scope of the present work, we can say that the gamma-ray bursts produced in this way will be long-duration, produce no afterglow and be should be correlated with the dark matter distribution. Note that the present data on gamma-ray bursts does not exclude such a correlation \cite{Tsvetkov:2001cc}.

\section{Conclusions}
We have shown that gauge-mediated Q-balls have an upper limit to their size due to the effects of gravity. Q-balls of this critical size will fragment into two daughter Q-balls of equal sizes. For those Q-balls that become trapped inside a neutron star, consumption of the entire star in an extremely rapid process. All such Q-balls are excluded by the observation of pulsars of age $10~\rm{Gyr}$.  An ultralight gravitino mass $m_{3/2} \sim 10~ \rm{eV}$ is required for Q-balls to be an acceptable dark matter candidate and avoid astrophage. This naturally avoids the usual gravitino problem and sets no bound on the reheating temperature. Moreover an ultralight gravitino may be found at the LHC~\cite{Hamaguchi:2007ji,Shirai:2009kn}. If Q-balls are sufficiently rare, their flux may not destroy all neutron stars and thereby account for some long-duration gamma-ray bursts. 

The author thanks Alexander Kusenko for very helpful discussions. 

\bibliography{qball}
\bibliographystyle{h-physrev3}

\end{document}